\documentclass[onecolumn]{emulateapj}

\usepackage{natbib,epsfig}
\usepackage{graphicx}
\usepackage{multirow}
\usepackage{amsmath}

\newcommand{\g}{\gamma}
\newcommand{\G}{\Gamma}
\newcommand{\eps}{\epsilon_{\rm acc}}
\newcommand{\plasma}{\rm pl}
\newcommand{\tg}{t_{\rm grow}}

\shorttitle{NAKAR ET AL.} \shortauthors{TWO-STREAM INSTABILITY IN
RELATIVISTIC SHOCKS}

\begin{document}

\title{Two-stream-like instability in dilute hot
relativistic beams and astrophysical relativistic shocks}
\author{Ehud Nakar \altaffilmark{1}, Antoine Bret \altaffilmark{2,3}, and
  Milo\v s Milosavljevi\'c \altaffilmark{4} }
\affil{1. Raymond and Beverly Sackler School of Physics \&
Astronomy, Tel Aviv University, Tel Aviv 69978, Israel\\
2. ETSI Industriales, Universidad de Castilla-La Mancha, 13071
Ciudad Real, Spain\\
3.Instituto de Investigaciones Energéticas, 13071 Ciudad Real,
Spain\\
4. Department of Astronomy and Texas Cosmology Center, University of
Texas, 1 University Station C1400, Austin, TX 78712, USA\\}

\begin{abstract}
Relativistic collisionless shocks are believed to be efficient
particle accelerators. Nonlinear outcome of the interaction of
accelerated particles that run ahead of the shock, the so-called
``precursor'', with the unperturbed plasma of the shock upstream, is
thought to facilitate additional acceleration of these particles and
to possibly modify the hydrodynamic structure of the shock. We
explore here the linear growth of kinetic modes appearing in the
precursor-upstream interaction in relativistic shocks propagating in
non and weakly magnetized plasmas: electrostatic two-stream parallel
mode and electrostatic oblique modes. The physics of the parallel
and oblique modes is similar, and thus, we refer to the entire
spectrum of electrostatic modes as ``two-stream-like.'' These modes
are of particular interest because they are the fastest growing
modes known in this type of system. Using a simplified distribution
function for a dilute ultra-relativistic beam that is
relativistically hot in its own rest frame, yet has momenta that are
narrowly collimated in the frame of the cold upstream plasma into
which it propagates, we identify the fastest growing mode in the
full $k$-space and calculate its growth rate. We consider all types
of plasma (pairs and ions-electrons) and beam (charged and
charge-neutral). We find that unstable electrostatic modes are
present in any type of plasma and for any shock parameters.  We
further find that two modes, one parallel ($k_\bot=0$) and the other
one oblique ($k_\bot \sim k_\|$), are competing for dominance and
that either one may dominate the growth rate in different regions of
the phase space. The dominant mode is determined mostly by the
perpendicular spread of the accelerated particle momenta in the
upstream frame, which reflects the shock Lorentz factor.  The
parallel mode becomes more dominant in shocks with lower Lorentz
factors (i.e., with larger momentum spreads). We briefly discuss
possible implications of our results for external shocks in
gamma-ray burst sources.
\end{abstract}


\section{Introduction}
Cosmic rays are believed to be accelerated in astrophysical
collisionless shocks. The leading acceleration process is diffusive
shock acceleration (DSA; e.g.,
\citealt{Bell78,Blandford1978,Blandford87}), where charged particles
are accelerated by crossing the shock back and forth. Thus, an
integral element of this picture is a precursor of accelerated
particles that runs ahead of the shock, and interacts with the
incoming, and mostly unperturbed, upstream plasma. The upstream
plasma must interact with the precursor in such a way as to deflect
the accelerated particles back into the shock and thus close the
Fermi cycle and enable farther acceleration.  It also provides an
early perturbation to the upstream plasma that may, through a
transfer of energy and momentum, and through an excitation of
electromagnetic perturbations in the plasma, modify the shock
structure \citep[e.g.,][]{Blandford80,Drury81}. If a different
 acceleration mechanism is at work (e.g., shock-drift acceleration;
\citealt{Webb83}), then the accelerated particles should similarly
cross the shock into the upstream, thus producing a
precursor, but these particles are not necessarily deflected
back into the shock.

The general picture described above should apply to both
relativistic and non-relativistic shocks and in various types of
upstream plasma and precursors, independelty of the plasma
magnetization level and composition. The details of the precursor
interaction, however, will depend on the specific system properties.
Here we explore a family of such systems that are thought to occur
in astrophysical environments: relativistic collisionless shocks
that propagate into unmagnetized or weakly magnetized plasma of
various compositions. Until recently, theoretical understanding of
particle acceleration in such shocks was based mostly on analytic or
semianalytic prescriptions \citep[e.g.,][and references
therein]{Achterberg01,Ellison02,Niemiec06}, in which the structure
of the medium that scatters the accelerated particles was assumed ad
hoc. A major breakthrough was achieved recently, when
two-dimensional first-principles particle-in-cell (PIC) simulations
of unmagnetized pair and ion-electron shocks provided direct
evidence for the acceleration of particles via DSA and for the
formation of a shock precursor
\citep{Spitkovsky2008a,Spitkovsky2008b,Keshet09,Martins09}.  As
expected, the numerical simulations show that a precursor of
accelerated particles runs ahead of the shock and interacts, by
exciting collective modes, with the initially undisturbed upstream
plasma. While opening a new window into the physical processes that
take place in these shocks, PIC simulations are still rather limited
in size and duration and an analytic treatment of the processes that
shape the shock is still required to extrapolate between the
simulations and real astrophysical systems. Moreover, existing
simulations have not converged to a steady state shock structure.
The maximum energy of the accelerated particles continues to grow
with the duration of the simulation and so does the spatial extent
of the precursor, as measured from the shock transition. The
properties of the leading edge of the precursor also keep evolving
in the simulations, e.g., the particle density at the precursor's
leading edge keeps dropping as the latter progresses away from the
shock transition. This implies that the kinetic instabilities that
will develop at the leading edge, on time and length scales that are
not sampled by existing simulations, may be very different than
those that are observed now in existing simulations.

There are several known instabilities that operate in
interpenetrating plasmas (see \citealt{Bret09} for an overview and
references). Three specific types of instabilities excited in shock
precursors could be important in relativistic unmagnetized or weakly
magnetized shocks. These are the kinetic two-stream-like
instability, the kinetic filamentation (or transverse Weibel)
instability, and the magnetohydrodynamic streaming instability
identified by \citet{Bell2004}. The Bell instability develops only
if the upstream plasma has some pre-existing seed magnetic field.
The filamentation instability is suppressed if the accelerated
particle precursor is dilute (i.e., its particle density is
sufficiently small compared to the upstream plasma density), and its
perpendicular momentum spread in the upstream frame is large enough
\citep{Silva2002}. The two-stream-like instability, which we define
to include the classical two-stream instability as well as its
oblique variant, is the only one that grows under any circumstances,
independent of the density and momentum spread in the upstream frame
of the shock precursor. Because of the two-stream-like instability's universal occurrence, we dedicate
this work to identifying its fastest growing modes and discussing
its potential role in relativistic unmagnetized shocks in
astrophysical and numerical settings.

The two-stream-like and filamentation instabilities (sometimes
referred to as ``Weibel'' instability
\citealt{Weibel,Fried1959,Moiseev63}) arise from the same branch of
the dispersion relation and compete with each other
\citep[e.g.,][]{Bret09,Michno10}. The two-stream parallel unstable
mode is found when the wave vector is aligned with the flow, while
the filamentation mode has its wave vector perpendicular to the
flow. For intermediate orientations, the so-called ``oblique'' modes
are equally unstable.  Since no magnetic field is generated, these
oblique modes are electrostatic, and are very similar to the
parallel two-stream mode, explaining why they can be labeled
``two-stream-like.'' For Maxwellian electron beam/plasma system in
the non-relativistic regime, two-stream modes grow faster than any
other unstable mode. In rather high density relativistic beams,
filamentation modes tend to dominate the linear phase. Oblique modes
become important in the dilute relativistic beam regime that is the
focus of the present work \citep{fainberg,BretPRL2008}.

Motivated by astrophysical considerations, we consider a
relativistic beam that is seen in the rest frame of the background
plasma to be having particle Lorentz factors narrowly distributed
around $\gamma$, and momenta that are directed within an angle of
$\sim \Gamma^{-1}$ from the beam axis,
where $\Gamma$ is the shock Lorentz factor. This is an ultra-relativistic beam that is
relativistically hot in its own rest frame, namely $\g \gg \G \gg
1$.  Unstable modes in this regime were explored very recently \citep{Lemoine11}, but
only for modes that satisfy $k_\|c=\omega_{\plasma}/\beta$ , where
$\omega_{\plasma}$ is the background plasma frequency and $\beta$ is
the velocity of beam particles. Here we explore the entire ($k_\|$,
$k_\bot$) space, including the pure parallel modes $k_\bot=0$, and
derive the fastest growing mode and its growth rate under the
assumption that the background plasma remains cold. Our results are
applicable to all types of unmagnetized shocks and precursors,
including beams containing ions and electrons, ions only, electrons
only, or pairs, and upstream plasmas containing ions and electrons or
pairs.

We do not go in the present work beyond the linear regime. The
nonlinear phase should be explored in the future (possibly with
numerical simulations), since it is on the detailed character of the
nonlinear development that a possible significant heating by, or
momentum transfer from, the accelerated particles, is contingent.
Similarly, the nonlinear development determines whether the
resulting electrostatic field can contribute accelerated particle
scattering in the shock upstream.

The paper is organized as follows. In \S 2 we discuss the
astrophysical context. In \S 3.1 we define our model distribution
function and derive the growth rates of the fastest growing modes.
In \S 3.2 we provide a simple physical description of the
instability. In \S 4 we discuss possible implications of our results
for GRB external shocks. In \S 5 we provide our conclusions.

\section{The astrophysical setting}\label{sec: astro setting}

Relativistic collisionless shocks are thought to occur in various
astrophysical environments, including in GRBs, pulsar wind nebulae,
active galactic nuclei, and microquasars. While the type of plasma
(ions and electrons or pairs) and the magnetization level is debated
in most of these systems, it is likely that astrophysical
relativistic shocks take place in both plasma types and over a large
range of magnetization levels. Among these environments, one of
the systems where the plasma properties are strongly constrained by
observations, is the external shock in GRBs. Here an ultra-relativistic
blast wave starts with very high Lorentz factors (up to
$\Gamma\sim1000$ and maybe even more) and decelerates with time to
Newtonian velocities. It propagates into a very weakly magnetized
circum-burst proton-electron plasma \citep[for reviews
see][]{Piran04,Meszaros06,Nakar07}. Another environment were the
plasma type is constrained is the termination shock of pulsar wind
nebulae, which is driven into a pair plasma (although some ions may
be present), but in this case, the magnetization level of the plasma
at the shock is debated \citep[for review see][]{Kirk09}.

Here, we consider collisionless relativistic shocks of both plasma
types, which are non-magnetized or only weakly magnetized. The
accelerated particles crossing into the shock upstream constitute a
beam of relativistic ions and/or electrons, or relativistic pairs.
The acceleration produces a spectrum of energies that is typically
approximated with a power-law,  where the largest number of the
accelerated particles, and most of the beam energy contained in the
accelerated particles, resides at the low-energy end of the
spectrum. Assuming that the particles are accelerated by DSA, the
ones that run ahead of the shock must be deflected by interacting
with the upstream plasma. Lower energy particles are deflected more
easily than those with higher energies. Therefore the particles with
higher energies can travel farther ahead of the shock and at any
distance from the shock the accelerated beam is dominated, both in
particle number and total energy, by the lowest energy particles
that reach that distance. These are also the particles that we
expect to be most susceptible to kinetic instabilities and are thus,
those that will dominate the growth rate of the two-stream-like
instability.\footnote{Our analysis validates this expectation.
Below, we find that the instability growth rate is slower for more
energetic particles (higher $\g$) and for beams with lower
densities. This result supports our monoenergetic approximation as
it implies that particles with much higher Lorentz factor than $\g$
are reacting to the the background plasma on much longer time scales
in the linear regime and can therefore be neglected.} Therefore, we
approximate the beam at a given distance from the shock as being
monoenergetic with particles Lorentz factor $\g$, as observed in the
upstream frame, which is the lowest Lorentz factor of beam particles
at that distance.  The rest frame of the beam, i.e., the frame in
which the beam is isotropic, moves with a Lorentz factor $\sim \G$
with respect to the upstream. The beam particles are also highly
relativistic in the beam rest frame, implying $\g/\G \gg 1$. Thus,
we work in the regime in which $\g \gg \G \gg 1$.

The charge neutrality of the beam depends on the specific setting. A
beam of pairs will always be charge-neutral, but a proton-electron
beam may be either charge-neutral or positively charged. If
electrons and protons are accelerated with similar efficiencies (as
numerical simulations suggest, see, e.g., \citealt{Spitkovsky2008a}) and
radiative losses can be neglected, then at any distance in front of
the shock the beam is charge-neutral and contains particles with
similar energies (i.e., $m_e\gamma_e\sim m_p\gamma_p$). Therefore,
since in the limit $\gamma_e,\g_p \gg 1$ the beam behavior depends
only on the product $m \gamma$, the beam is similar to a beam of
pairs. If, on the other hand, electrons are not accelerated as
efficiently as protons, either due to less efficient injection into
the DSA cycle, or due to higher radiative losses, the accelerated
particle beam is positively charged. Note that the electrons and
protons of an initially cold upstream always behave differently.

An instability can have a strong impact on the structure of an
astrophysical system if its amplitude becomes nonlinear. In order to
do so it must have enough time to grow, namely, the time that the
instability has to grow, $\tg$, must be significantly longer than
the instability growth time, given by the inverse growth rate
$1/\delta$. In the context of our setup, the value of $\tg$ for an
upstream fluid element is the time between the first encounter of
the fluid element with the accelerated particle precursor and the
time that this element is swept up by the shock transition. If the
shock is spherical and its radius is $R$, then $\tg \lesssim
R/c\G^2$, where $\tg$ is measured in the upstream rest frame. In a
planar geometry, which is typically used in numerical simulations,
$\tg \leq t/\G^2$, where $t$ is the age of the shock (both times are
measured in the upstream frame). Note that a beam particle spends a
time much longer than $\tg$ in the shock upstream. In spherical
geometry it can be as long as $\sim R/c$ and in planar geometry it
can be as long as $t$. Therefore, beam particles have much more time
to be affected by the interaction with the upstream.

\section{The fastest growing mode}

\subsection{The dispersion relation}
\label{sec:dispersion_relation}

Consider a dilute monoenergetic particle beam that
propagates in a dense cold plasma. The rest mass of each beam
particle is $m_{\rm b}$ and its energy is $m_{\rm b} \gamma$, as
measured in the background plasma frame.\footnote{If the beam
contains several particle species, then we assume energy
equipartition between particles of different masses. For example, in
a proton-electron beam $m_e \gamma_{{\rm b},e}=m_p \gamma_{{\rm
b},p}\,(\equiv m_{\rm b}\gamma)$. Given the highly relativistic
setting considered here, the slight velocity drift between species
of different mass (e.g., since $\gamma_{{\rm b},e}\neq\gamma_{{\rm
b},p}$) can be neglected.} In this frame the beam momenta are narrowly
collimated within a half opening angle $\sim \Gamma^{-1}$, where $\g \gg \G
\gg 1$. The following calculation makes no assumptions about the
beam and plasma composition.

Since we consider a monoenergetic beam, we can neglect the
longitudinal momentum spread of the beam, as seen in the upstream
frame, in which the calculations will be conducted. The transverse
momentum spread in this frame is $\sim m_{\rm b}\gamma/\Gamma$. In
order to investigate the stability of such a beam, we consider a
system containing a beam of density $n_{\rm b}$ and mean velocity
$v_{\rm b} \approx c$ that interpenetrates a cold background plasma
of density $n_{\plasma} \gg n_{\rm b}$.  Both densities are measured
in the upstream plasma rest frame (in the shock frame the beam
density is smaller by a factor of $\G$ while the upstream plasma
density is larger by the same factor). The beam density accounts for
all species (e.g., in a proton-electron beam, we have $n_{\rm
b}=n_{{\rm b},e}+n_{{\rm b},p}$), while the plasma density includes
only the lightest species (e.g., in a proton-electron plasma, we
have $n_{\plasma}=n_{{\plasma},e}$ and in a pair plasma, we have
$n_{\plasma}=n_{{\plasma},e^-}+n_{{\plasma},e^+}$), since the
heavier species is assumed to be static. The system is charge
neutral so plasma drift neutralizes the beam charge. In a
proton-electron plasma and beam, the electrons drift with velocity
$v_{\plasma} \approx C (n_{\rm b}/n_{\plasma}) v_{\rm b}$ and cancel
the beam current, where
\begin{equation}
C\equiv \frac{n_{{\rm b},p}-n_{{\rm b},e}}{n_{\rm b}}
\end{equation}
is the fractional beam
charge satisfying $-1\leq C\leq 1$ . We show later that this drift has a negligible effect on the
instability spectrum and on the growth rate.

Given the symmetry of the dispersion relation with respect to the
rotation around the parallel direction, a two-dimensional survey of
the $k$-space is required \citep{BretPRL2005}. In general, such
calculation requires solving the linearized Maxwell-Vlasov
equations. The electrostatic, or longitudinal, approximation, i.e.,
$\mathbf{k}\times \mathbf{E}=0$, is not appropriate in general
because the modes evolve continuously from $\mathbf{k}\times
\mathbf{E}=0$ along the beam  direction, to $\mathbf{k}\times
\mathbf{E}\neq 0$ in the perpendicular direction. For example, the
filamentation instability is found for $\mathbf{k}\perp
\mathbf{v}_{\rm b}$ and has $\mathbf{k}\times \mathbf{E}\neq 0$, but
it need not have $\mathbf{k}\cdot \mathbf{E}= 0$
\citep{BretFilaLongi}. Thus, in general, the electrostatic
approximation can lead to a neglect of the most unstable modes.
However, in the specific case of interest, where the beam is
relativistic and dilute, oblique or parallel parts of the
spectrum govern the linear phase of the instability
\citep{BretPRL2008} and the electrostatic approximation is valid
\citep{BretPRE2004,GremilletPoP2006}. Thus, in
what follows, we adopt the electrostatic approximation.

The longitudinal dispersion equation
thus reads \citep{Ichimaru},
\begin{equation}\label{eq:disper}
    \varepsilon(\mathbf{k},\omega)=1+\sum_\sigma\frac{4\pi n_\sigma
      e^2}{k^2}\int\frac{\mathbf{k}\cdot \nabla_{\mathbf p} f_\sigma(\mathbf{p})}{\omega-\mathbf{k}\cdot \mathbf{v}}d^3p=0,
\end{equation}
where the sum runs over the species $\sigma$ which are here the beam
species, plus the lightest species of the cold background plasma. We
now choose the $z$ axis as the direction of beam propagation, and
label $x$ the normal direction. The momentum distribution function
of each beam species reads:
\begin{equation}\label{eq:fbeam}
    f_{\rm b}({\mathbf p}) = \frac{1}{2P_T}\delta(p_z-\gamma m_{\rm b} v_{\rm
      b})\left[\Theta(p_x+P_T)-\Theta(p_x-P_T)\right],
\end{equation}
where $P_T\equiv \gamma m_e v_{\rm b}/\Gamma$, while the
distribution function of each (background) plasma species,  is
\begin{equation}\label{eq:fplasma}
    f_{\plasma}({\mathbf p})= \delta(p_z + C m_{\plasma} v_{\plasma})\delta(p_x),
\end{equation}
where $m_{\plasma}$ is the mass of the lightest plasma particle,
typically an electron, and $\Theta(t)$ is the Heaviside step
function. Note that $\gamma m_{\rm b}$ is equal for all beam species
and that in a pair plasma $C=0$.

The dispersion equation obtained when inserting these functions in
equation (\ref{eq:disper}) can be written concisely as
\begin{equation}\label{eq:disperOK}
0 = 1 - \frac{1}{(\tilde{\omega}-C Z_z \alpha )^2}-\alpha\frac{
m_{\plasma}}{m_{\rm b}} \Psi ,
\end{equation}
where
\begin{equation}\label{eq:Psi}
\Psi \equiv\frac{Z_x^2}{Z_x^2+Z_z^2}\frac{
\phi(\rho)/\gamma}{\left[Z_z -\tilde{\omega} \phi(\rho)\right]^2 -
Z_x^2\rho^2}
+\frac{Z_z}{2\gamma^3\left(Z_x^2+Z_z^2\right)}\int_{-1}^1
\frac{Z_z-\rho t  \gamma ^2   (Z_x-\rho t Z_z )}{\phi(\rho t)
\left[Z_z  +\rho t Z_x    -\tilde{\omega} \phi(\rho t)\right]^2} dt.
\end{equation}
Here $\g$ is the Lorentz factor of beam particles with mass $m_{\rm
  b}$ and we define the following symbols
\begin{equation}\label{eq:var}
    \alpha\equiv \frac{n_{\rm b}}{n_{\plasma}},~~\mathbf{Z}\equiv\frac{\mathbf{k} c}{\omega_{\plasma}},~~\rho\equiv\frac{1}{\Gamma},~~\tilde{\omega}\equiv\frac{\omega}{\omega_{\plasma}},~~\phi(\tau)\equiv\frac{1}{\gamma}\sqrt{1+\gamma^2(1+\tau^2)},
\end{equation}
where  $\omega_{\plasma}^2=4\pi n_{\plasma} e^2/m_e$ is the plasma
frequency of the background lightest particles and we approximate
$v_{\rm b}\approx c$. The second term on the r.h.s.\ of equation
(\ref{eq:disperOK}) pertains to the lightest particle species of the
background plasma, and the last term accounts for the beam response.

Calculations can be simplified in the present regime noting that
$0<\rho\ll 1$ and $\g \gg 1$, which implies that $\phi(\rho)\sim 1$ and
$\phi(\rho t)\sim 1$ for $-1\leq t\leq 1$. The quadrature in
equation (\ref{eq:Psi}) can be further simplified to obtain
\begin{equation}\label{eq:Psi1}
\Psi \approx  \frac{1}{\gamma(Z_x^2+Z_z^2)\left[(Z_z
-\tilde{\omega})^2 - Z_x^2\rho^2\right]}
\left[Z_x^2+\frac{Z_z^3}{6\gamma^2}\frac{(3+\gamma^2\rho^2)}{(Z_x^2+Z_z^2)}\right].
\end{equation}

All types of beams and plasmas exhibit a similar unstable spectrum. The
reason is that the only dimensionless parameter that cannot be removed
through scaling is the fractional beam charge $C$. It is the beam charge density that determines whether the
background plasma drifts and what is the direction of the drift. However, the drift velocity is proportional
to the density ratio $\alpha$ and is thus very small.
Therefore in equation (\ref{eq:disperOK}), the parameter $C$ is only
found in a product with the
density ratio $\alpha$ and the effect of space charge in the beam is negligible (note that the value of the third factor in the
product is $Z_z\sim 1$ for the most unstable mode).

\begin{figure}[t]
\begin{center}
 \includegraphics[width=0.5\textwidth]{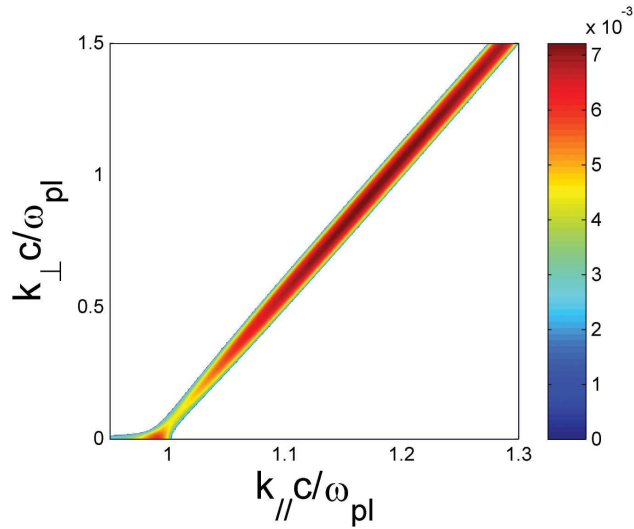}
\end{center}
\caption{Growth rate in units of the plasma frequency as a function
of the dimensionless wave vector
$\mathbf{Z}=\mathbf{k}c/\omega_{\plasma}$, in the case of a pure
electron beam. Parameters are $\alpha=10^{-2}$, $\gamma=100$, and
$\Gamma=5$.} \label{fig1}
\end{figure}

\begin{figure}[t]
\begin{center}
\includegraphics[width=0.5\textwidth]{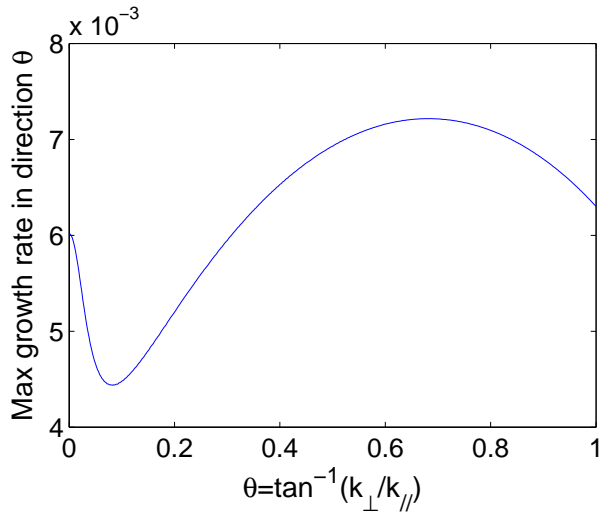}
\end{center}
\caption{Growth rate in units of the plasma frequency as a
function of the mode angle $\tan^{-1} (k_\perp/
k_\parallel)$. Parameters are the same as in Figure \ref{fig1}.}
\label{fig2}
\end{figure}

A typical two-dimensional map of the unstable spectrum for a pure
electron beam is shown in Figure \ref{fig1}, for $\alpha=10^{-2}$,
$\gamma=100$, and $\Gamma=5$. The growth rate of the fastest growing
unstable mode as a function of the mode angle $\tan^{-1} (k_\perp/
k_\parallel)$ is shown in Figure \ref{fig2}. The spectrum exhibits
two local maxima. One corresponds to the parallel mode with
$(Z_\parallel,Z_\perp) \approx (1,0)$ and the other to the oblique
mode with $(Z_\parallel,Z_\perp) \approx (1.2,1)$. The relative
height of the two maxima, and hence the character of the fastest
growing mode---whether it is parallel or oblique---depends on the
specific values of the parameters. Also noticeable is the narrowness
of the range of wave vectors containing unstable modes. With a high
beam Lorentz factor, a mode can only grow out of its interaction
with the beam if a certain resonance condition is strictly fulfilled
(we discuss this condition in \S\,\ref{sec: intuitive description}).
Modes which are ``out of tune'' are quickly damped away.

The total absence of unstable modes in the perpendicular direction
is not always just a direct consequence of the electrostatic
approximation, which does not account for such non-longitudinal
modes. The theory of unstable filamentation modes for the present
setting shows that they are completely shut down if the transverse
beam angular spread as seen in the upstream frame $\sim
\Gamma^{-1}$ is larger than $[\alpha m_{\plasma}/(\gamma m_{\rm
b})]^{1/2}$, that is, when the density ratio is limited by
\citep{Silva2002,BretPRE2005,Rabinak10}
\begin{equation}
\alpha<\frac{\gamma m_{\rm b}}{\Gamma^2 m_{\plasma}}\ \ \
(\textrm{filamentation suppression}) .
\end{equation}
Since we work here in the regime in which $\alpha\ll
1\ll\Gamma\ll\gamma$, filamentation modes are unstable only in a
small portion of the phase space, especially in an ion-electron
plasma. Moreover, if the first shock-crossing cycle of the DSA
accelerates particles to Lorentz factors $\g \sim
\G^2$ \citep[e.g.,][]{Gallant99}, then the filamentation modes are
stable in the entire precursor for any type of plasma. Finally, let
us emphasize that even when filamentation modes are unstable, their
growth rate is slower than the fastest growing two-stream mode that
we find here, so they do not dominate anyway.

The growth rate of the fastest purely parallel growing mode can be
found analytically. The dispersion equation for such wave vectors,
$\mathbf{Z}=Z_z \hat{\mathbf z}$, is easily derived from equations
(\ref{eq:disper}) and (\ref{eq:Psi}) and reads
\begin{equation}\label{eq:disp par}
    0=1-\frac{1}{(\tilde{\omega}-C Z_z \alpha )^2}-\alpha\frac{m_{\plasma}}{m_{\rm b}}\frac{ \left(3+  \gamma^2 \rho ^2\right)}{3 (\tilde{\omega}-Z_z)^2 \gamma ^3}.
\end{equation}
This relation is mathematically similar to the dispersion relation of the two-stream
instability. The maximum parallel growth rate $\delta_{\rm par}$ is
reached for $Z_z\sim 1$ and reads
\begin{equation}\label{eq:GR_par}
  \delta_{\rm par}
    \approx \frac{3^{1/6}}{2^{4/3}}\left(\frac{m_{\plasma}}{m_{\rm b}}\frac{ \alpha }{\gamma  \Gamma^2}\right)^{1/3}\omega_{\plasma}.
\end{equation}

The growth rate of the fastest growing oblique mode can be solved for
numerically. We conducted a systematic numerical search for the
fastest growing mode for representative choices of $\alpha$,
$\gamma$, and $\Gamma$. We find that when $\alpha\ll 1$ and
$1\ll\Gamma\ll\gamma$, the maximum oblique growth rate $\delta_{\rm
obl}$ reads
\begin{equation}\label{eq:GR oblique}
   \delta_{\rm obl}\approx \frac{1}{3}\left(\frac{m_{\plasma}}{m_{\rm b}}\frac{\alpha\Gamma}{ \gamma}\right)^{1/2} \omega_{\plasma} .
\end{equation}

Equating the two growth rates yields the criterion for the dominance
of each, so that the global fastest growing mode grows at the rate
\begin{equation}\label{eq: delta max}
\delta_{\rm max}=\left\{\begin{array}{lr}
                          \delta_{\rm par} ~~~~;~~~~& \frac{2^{8/7}}{3}
\left(\frac{m_{\plasma} \alpha}{m_{\rm b} \gamma}\right)^{1/7} \G < 1
, \\
&\\
                          \delta_{\rm obl} ~~~~;~~~~& \frac{2^{8/7}}{3}
\left(\frac{m_{\plasma} \alpha}{m_{\rm b}
\gamma}\right)^{1/7}\G > 1 .
                        \end{array} \right.
\end{equation}
Some of the general trends evident in equations (\ref{eq:GR_par}) and
(\ref{eq:GR oblique}) were to be expected. Namely, the growth rate
weakens when the beam density decreases or when the inertia of the beam
particles increases. The fact that the growth rate has two local
maxima with opposite dependencies on $\G$ is nontrivial and is
discussed in the next section. It is interesting to compare this
result to the scenario in which a cold beam interacts with a cold
plasma \citep{fainberg}.  There, the fastest growing mode is always
oblique and its growth rate scales as $(\alpha/\gamma)^{1/3}$.

Finally, we stress that equations  (\ref{eq:GR_par}) and (\ref{eq:GR
oblique}) provide the growth rates for monoenergetic beam with an
idealized distribution function given in equation (\ref{eq:fbeam}). The true
distribution function in collisionless shocks with DSA is of course
more complicated. Since the exact value of the growth rate and its
dependence on the various parameters depend on the exact shape of
the distribution function, we expect the result to be somewhat
different if a more realistic distribution is considered. However,
we still expect that the key proportionalities evident in equations
(\ref{eq:GR_par}) and (\ref{eq:GR oblique}) persist in the general
case, although the exact values of the power-law indices may vary.
We also expect that there will still be a competition between the
parallel and oblique modes.

\begin{figure}
\begin{center}
\includegraphics[width=0.5\textwidth]{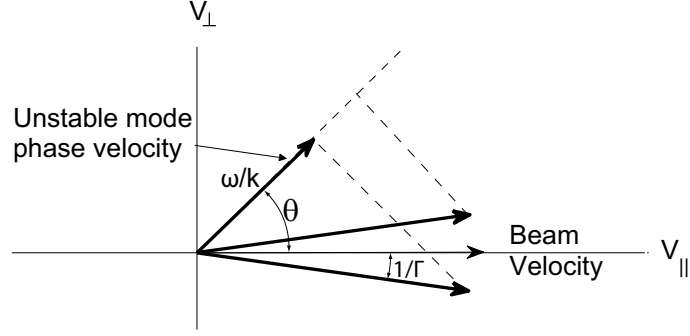}
\end{center}
\caption{Illustration of the beam and wave mode geometry.}
\label{fig3}
\end{figure}

\begin{figure}[t]
\begin{center}
\includegraphics[width=0.5\textwidth]{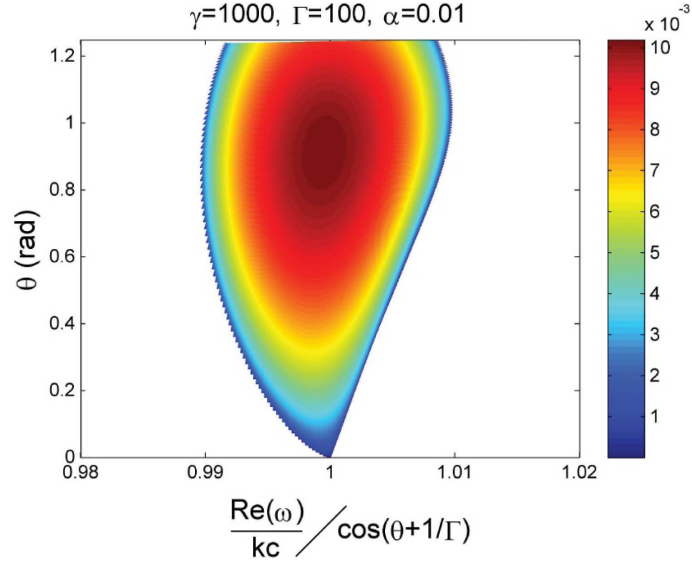}
\end{center}
\caption{Growth rate map in the plane defined by $\frac{{\rm
Re}(\omega)}{ck}[\cos (\theta+\Gamma^{-1})]^{-1}$ and $\theta$ for
unstable modes with $0<Z_\parallel,Z_\perp<2$. The color shows the
growth rate. Parameters are $\alpha=10^{-2}$, $\gamma=1000$, and
$\Gamma=100$. The narrow concentration of the unstable modes around
the resonant condition $\frac{{\rm Re}(\omega)}{ck}[\cos
(\theta+\Gamma^{-1})]^{-1}=1$ is clearly evident.} \label{fig4}
\end{figure}

\subsection{Intuitive physical description}\label{sec: intuitive
description}

The oblique mode is a two-stream-like electrostatic mode, and as
such it can be understood with the aid of the intuition gained from
the consideration of a one-dimensional, longitudinal two-stream
mode. In one dimension, a wave mode interacts with near-resonant
particles that move with velocities similar to the phase velocity of
the mode. These particles ``surf" the wave in the sense that they
spend a long time in phase with the wave. During surfing, the
velocity of near-resonant particles further converges to the phase
velocity of the wave. The particles that are slightly faster than
the wave tend to transfer energy to the wave, and thus excite it,
while those that are slightly slower than the wave tend to receive
energy from it, and thus damp the wave. Since growing modes are
those with a net energy gain, a mode is unstable if the number of
particles moving slightly faster than the wave exceeds that of those
moving slightly slower. Thus a distribution function $f(v)$ should
be two-stream unstable only if $\left(\partial f/\partial
|v|\right)_{v_{\rm phase}}
> 0$, where $v_{\rm phase}={\rm Re}(\omega)/k$ is the phase
velocity of the wave mode.

The three dimensional case is similar, except that now it is
$v_{\mathbf k}$, the component of the velocity parallel to the wave
vector $\mathbf{k}$, that determines the stability of the mode. In
the configuration under consideration all the accelerated particles
move at about the speed of light and  have momenta that are
collimated to within the small angle $\sim\Gamma^{-1}$ from the
direction of shock propagation (see Figure \ref{fig3}). For any
generic wave vector, except for the almost perpendicular ones, the
range of particle velocities projected onto the wave vector does not
include the zero velocity. Thus a mode with phase velocity equal to
the minimum of the projected beam particle velocities has many
particles that are moving slightly faster than the wave but none
that are moving slower than the wave.  Based on the one-dimensional
analysis in the preceding paragraph, the mode is unstable.
Neighboring modes with an excess of particles moving faster than the
wave should also be unstable. All unstable two-stream oblique modes
satisfy the resonance criterion
\begin{equation}
\frac{v_{\rm phase}}{c}\equiv \frac{{\rm Re}(\omega)}{c k} \approx
\cos (\theta+\Gamma^{-1}) ,
\end{equation}
where $\theta$ is the angle between $\mathbf{k}$ and the $z$ axis.
This include the parallel mode, $\theta=0$, and oblique modes up to
$\theta=\pi/2-\Gamma^{-1}$. Figure \ref{fig4} shows a growth rate
map in the plane of  $\frac{{\rm Re}(\omega)}{ck}[\cos
(\theta+\Gamma^{-1})]^{-1}$ and $\theta$, where the resonance is
clearly evident. The resonance condition for instability in the
regime that we consider, up to terms of the order of $\G^{-2}$, can
therefore be written as
\begin{equation}
\label{eq:resonance_condition} {\rm Re}(\omega)\approx
ck_\|-\frac{ck_\bot}{\G} .
\end{equation}

This intuitive description does  not select the fastest growing mode
among all the unstable modes. In \S \ref{sec:dispersion_relation} we
analyzed the dispersion relation to pin down the dominant mode and
derived the dependence of growth rate on the shock Lorentz factor
$\Gamma$ in the parallel and oblique regimes.  The different dependencies
of the growth rates on $\G$ can be understood as follows. In
general, the growth rate depends on two competing and opposite
factors. One is  the number of particles that are in resonance with
the mode. The energy of these particles can in principle be tapped
toward contributing to the growth of the mode. The other relates to
efficiency with which energy transfer from the particles to the mode
can take place. The second factor is the reason that oblique modes
play an important role in relativistic beams. It is much easier the
reduce the projected velocity of an ultra-relativistic particle in
the direction of $\mathbf{k}$, and thus to trap the particle in a
resonance with the mode and transfer its energy to the mode, by
deflecting the particle sideways, than by decelerating the particle.
From this it follows that larger value of the angle between the
resonant particles and the wave, $\theta+\G^{-1}$, contribute to
increasing the growth rates. On the other hand, now returning
to the first aforementioned factor, larger values of $\G$ and
smaller values of $\theta$ imply a distribution function that is
more compact when projected onto the direction of $\mathbf{k}$. This
ensures that a larger number of nearly-resonant particles is
available. From this, however, it follows that having a larger value of
$\theta+\G^{-1}$ also works in the opposite direction, namely,
toward decreasing the growth rates. It is the balance between these to
opposite dependencies of the two effects on $\theta+\G^{-1}$ that
determines the location in $\mathbf{k}$ space of the fastest growing
mode. Now, in oblique modes, where $\theta \gg \G^{-1}$ and thus
$\theta+\G^{-1}\approx \theta$, increasing $\G$ only increases the
compactness of the distribution function in the direction of
$\mathbf{k}$, thereby increasing the growth rate of the fastest
growing oblique mode. In parallel modes where $\theta\approx 0$ and
thus $\theta+\G^{-1}\approx \Gamma^{-1}$, increasing $\G$ increases
the number of resonant beam particles but makes it harder to exploit
their energy. It turns out the latter factor wins and the growth
rate increases when $\G$ is smaller.

\section{Application to external shocks in GRBs}

Here we assess under which conditions the two-stream-like
instability in the precursor of GRB external shocks can become
nonlinear and strongly affect the shock structure, assuming that the
acceleration process is DSA. We do so by comparing $\tg$ in such a
shock (see the definition in \S\,\ref{sec: astro setting}) to
$1/\delta_{\rm max}$ found in \S\,\ref{sec:dispersion_relation}. A
GRB external shock is a spherical relativistic collisionless shock
that propagates in a weakly magnetized proton-electron plasma.
Electron acceleration in GRB external shocks is expected to be
limited by the radiative cooling \citep{Li06,Couch08} and thus the
precursor beam should contain mostly protons at higher energies.
The accelerated particle spectrum is taken to be a power-law
$dN_{\rm acc}/d\gamma \propto \gamma^{-p}$ where $N_{\rm acc}$ is
the total number of accelerated particles in the shocked region and
the power law index is in the range $2<p<3$. We assume that the
minimum accelerated particle energy is $\sim \G^2$, where $\Gamma$
is the shock Lorentz factor, as expected for the accelerated
particles that have completed their first full Fermi cycle.  The
density of the accelerated particle precursor can then be
parameterized by the fraction $\eps$  of the shock energy that goes
into the accelerated particles. The total accelerated particle
number is then $N_{\rm acc} \sim  \eps E /(\G^2 m_p c^2)$, where $E$
is the total shock energy. The density of accelerated particles with
Lorentz factors $\sim \gamma$, as measured in the upstream frame, is
$n_{\rm acc} (\g) \sim (\G^2/R^3) (\g/\G^2)^{1-p} N_{\rm acc}$,
where $R$ is the shock radius. We assume that the density $n$ of the
unshocked upstream is constant. Hydrodynamics of an
ultra-relativistic blast wave \citep[e.g.,][]{BlandfordMcKee76}
relates $n$ to the blast wave energy $E$, radius $R$, and Lorentz
factor $\G$; with this, one finds that the density ratio measured in
the upstream frame, $\alpha\equiv n_{\rm acc}/n$, is
\begin{equation}\label{eq:GRB_alpha}
    \alpha(\g) \sim 0.1
    \left(\frac{\epsilon_{\rm
        acc}}{0.1}\right)\left(\frac{\G}{100}\right)^{2p}\left(\frac{\g}{10^6}\right)^{1-p} .
\end{equation}
Note that $\alpha$ retains no explicit dependence on $E$ or $n$. At
any distance from the shock the accelerated particle density is
dominated by the lowest energy particles that reach that distance.

We first determine which among the oblique and parallel modes
dominates. Substituting equation (\ref{eq:GRB_alpha}) into (\ref{eq:
delta max}), using $m_{\rm b}=m_p$, we find that with
$\gamma\sim10^6$ (the dependence on the particle Lorentz factor is
weak), for $\G \gtrsim 60$ the oblique mode dominates while for $\G
\ll 60$ the parallel mode dominates. Since most of the afterglow
observations cover the evolution when $\G < 60$, in what follows, we
will focus on the low-$\Gamma$ regime in which the parallel mode
dominates. The growth rate derived in equation (\ref{eq:GR_par})
decreases with increasing $\g$ and scales as $\delta_{\rm max}
\propto (\alpha/\g\G^2)^{1/3} \propto \g^{-p/3} \G^{2(p-1)/3}$. The
time available for growth is $\tg \sim \Delta(\gamma)/c$ where
$\Delta(\gamma)$ is the distance from the shock that particles with
Lorentz factors $\gamma$ can reach. The maximum distance is
$\Delta_{\rm max} \sim R/(10\G^2)$, and it is reached by the protons
that are accelerated to the highest energies $\g_{\rm max}$. A
pre-existing shock upstream magnetic field with a strength of a few
$\mu$G and with a coherence length that is larger than the shock
radius, as expected in the ISM, can accelerate protons up to
$\g_{\rm max}\sim 10^6$. If accelerated particle streaming in the
shock precursor can somehow amplify the pre-existing field to the
point of equipartition with the rest energy of the circum-burst
medium, proton acceleration to energies $\g_{\rm max}\sim 10^{10}$
may be possible \citep{Milos06b}. We approximate the dependence of
the distance from the shock on the particle Lorentz factor with a
power-law $\Delta(\gamma)\propto \g^s$. If all the accelerated
particles are deflected back into the shock by a magnetic field with
coherence length larger than $R$, then we  have $s=1$. If, on the
other hand, they are deflected by a small-scale with coherence
length $\lambda_B\ll R$, then we have $s=2$ at all distances
$\Delta>\lambda_B$.

The maximum number of $e$-foldings by which an unstable mode can
grow is the ratio $\tg/\delta_{\rm max}^{-1}$, implying that
$N_{\text{ e-fold}}(\g) = \delta_{\rm max}(\g) \Delta(\g)/c $. Using
equations (\ref{eq:GR_par})  and (\ref{eq:GRB_alpha}), we find
\begin{equation}
N_{\text{e-fold}}(\g) \approx 100
    \left(\frac{\epsilon_{\rm acc}}{0.1}\right)^{1/3}
    \left(\frac{\G}{100}\right)^{(2p-10)/3}
    \left(\frac{n}{1\,\textrm{cm}^{-3}}\right)^{1/6}
    \left(\frac{E}{10^{53}\,\textrm{erg}}\right)^{1/3}
    \left(\frac{\g_{\rm max}}{10^6}\right)^{-p/3}
    \left(\frac{\g}{\g_{\rm max}}\right)^{s-p/3} .
\end{equation}
The dependence on $E$ and $n$ is rather weak, while the scaling with
$\G^{-1}$ varies from being linear to being quadratic for $2<p<3$.
This implies that if the accelerated protons carry a non-negligible
fraction of the total energy of the shock, e.g., $\epsilon_{\rm acc}
\gtrsim 10^{-3}$, but the maximum proton energy is relatively low,
$\g_{\rm max} \lesssim 10^6$, then as long as the external shock
remains relativistic  $N_{\text{ e-fold}}(\g_{\rm max}) \gg 1$, and
at the largest distance from the shock there is enough time for the
two-stream modes to become nonlinear, then the instability can
strongly influence the state of the upstream plasma and accelerated
particle beam dynamics. If, in the other extreme, $\g_{\rm max} \sim
10^{10}$  and $s=1$, then $N_{\text{e-fold}}(\g_{\rm max}) \lesssim
1$ throughout the external shock, and the two-stream instability
should not affect its structure. For intermediate values of $\g_{\rm
max}$ or when $s=2$, the number of $e$-foldings is below unity when
$\G$ is large, but it can become larger than unity as $\G$
decreases; this transition may imply a change in the shock
structure. Finally, we note that for low values of $\g$, i.e., close
to the shock, and for high values $\G$, it is possible that the
accelerated particle density exceeds the upstream plasma density,
$\alpha>1$, in which case our analysis is no longer valid.

\section{Summary}

We explored the linear growth rate of electrostatic two-stream-like
modes that grow when a dilute ultra-relativistic beam, which is
relativistically hot in its own rest frame, propagates in a cold,
unmagnetized or weakly magnetized, background plasma. This
beam-plasma configuration describes the precursor containing
accelerated particles that runs ahead of an ultra-relativistic,
unmagnetized or weakly magnetized, collisionless shock. We
considered various types of plasma (electrons and ions or pairs) and
beams of various compositions and studied the fastest growing
kinetic modes. Such modes may thus be the principal agent of
nonlinear precursor-upstream interaction that is expected in these
shocks and that is also apparent in numerical simulations
\citep[e.g.,][]{Keshet09}.

Our main finding is that for any beam and plasma composition and for
any shock and beam parameters, as long as $\g \gg \G \gg 1$ and the
background plasma is cold, there exist two-stream-like unstable modes.
These are the modes that satisfy the resonance criterion ${\rm
Re}(\omega)/c k \approx \cos (\theta+\Gamma^{-1})$, where $\theta$
is the angle between wave vector $\mathbf{k}$ and the direction of
beam propagation. This includes the parallel mode with
$\theta=0$, and oblique modes up to $\theta=\pi/2-\Gamma^{-1}$. The
resonance criterion can also be written as ${\rm Re}(\omega)\approx
ck_\|-ck_\bot/\G$. This behavior departs from that of the
filamentation (or transverse Weibel) modes, which are suppressed
over a large range of the parameter space. Moreover, in the
interaction of a dilute accelerated particle precursor with a
dense, cold, unmagnetized, and unperturbed upstream, the fastest
growing electrostatic two-stream-like mode always grows faster than
any of the filamentation modes when the latter are unstable. We
therefore do not expect filamentation modes to play a role in the
shock precursor containing accelerated particles, but possibly only
in the shock transition itself.

We find that the spectrum of unstable electrostatic modes exhibits
two local maxima. One contains parallel modes, with wave vectors
$(k_\parallel,k_\perp) \sim \frac{\omega_{\plasma}}{c}(1,0)$, and
the other contains oblique ones with $(k_\parallel,k_\perp)\sim
\frac{\omega_{\plasma}}{c}(1+\G^{-1},1)$. The relative height of the
two maxima, and therefore the character of the dominant
mode---whether it is of the parallel or oblique type---depends on
the parameters and is most sensitive to $\G$. For large values of
$\G$, an oblique mode dominates, while for small values of $\G$, it
is the parallel mode that dominates. The criterion for finding the
dominant mode exhibits a similar trend with $m_{\plasma}
\alpha/(m_{\rm b} \g)$, but this dependence is much weaker.
Nevertheless, in pair or electron beams, the oblique mode dominates
over a larger portion of the parameter space than it does in ion
beams.

The time that is available to an unstable mode for its growth  is
limited in the precursor of astrophysical ultra-relativistic shocks.
Nevertheless, if sufficient time is available for the fastest
growing mode to become nonlinear, the nonlinear outcome of the
instability may significantly affect the shock structure. It can
lead to momentum transfer from the accelerated particles to the
shock upstream, thereby modifying the hydrodynamic structure of the
shock. We compared the growth time of the fastest growing
two-stream-like mode to the time available for its growth in GRB
external shocks. We find that for all except for the earliest
phases, when the shock Lorentz factor has dropped below $\sim 100$,
the parallel mode dominates. If the maximum Lorentz factor of the
accelerated protons is modest, $\g_{\rm max} \sim 10^6$, then this
mode has enough time to become nonlinear, while if $\g_{\rm max}$ is
significantly larger, then the mode may become nonlinear only for
relatively low values of $\G$. We conclude that the electrostatic
two-stream-like instability may play an important role in
accelerating particles and in shaping of the structure of the
external shocks in GRBs.


We would like to thank Uri Keshet and Anatoly Spitkovsky for helpful
discussions. E.\,N.\ was supported in part by the Israel Science
Foundation (grant No.\ 174/08) and by an EU International
Reintegration Grant. A.\,B.\ was supported by projects ENE2009-09276 of
the Spanish Ministerio de Educaci\'{o}n y Ciencia and
PEII11-0056-1890 of the Consejer\'{i}a de Educaci\'{o}n y Ciencia de
la Junta de Comunidades de Castilla-La Mancha.


\end{document}